\documentclass[12pt]{article}
\usepackage{fullpage}
\usepackage{graphicx}
\newcommand{\be}{\begin{equation}}
\newcommand{\ee}{\end{equation}}

\renewcommand{\hat}{\widehat}
\renewcommand{\tilde}{\widetilde}

\font\mybb=msbm10 at 11pt

\def\bb#1{\hbox{\mybb#1}}

\def\bR {\bb{R}}

\newcommand{\news}{\setcounter{equation}{0}\quad}
\def\ben{\begin{equation}}
\def\een{\end{equation}}
\def\bea{\begin{eqnarray}}
\def\eea{\end{eqnarray}}
\begin{document}
\title{
\vskip 2cm Domain Walls and Double Bubbles}
\author{Mike Gillard and Paul Sutcliffe\\[10pt]
{\em \normalsize Department of Mathematical Sciences,
Durham University, Durham DH1 3LE, U.K.}\\[10pt]
{\normalsize Email:\  mike.gillard@durham.ac.uk,
 \ p.m.sutcliffe@durham.ac.uk}
}
\date{March 2009}
\maketitle
\begin{abstract}
We study configurations of intersecting domain walls in a Wess-Zumino
model with three vacua. We introduce a volume-preserving flow and show
that its static solutions are configurations of intersecting domain walls
that form double bubbles, that is, minimal area surfaces which enclose
and separate two prescribed volumes. To illustrate this field
theory approach to double bubbles,
we use domain walls to reconstruct the phase diagram for double bubbles
 in the flat square two-torus and also construct all known examples of 
double bubbles in the flat cubic three-torus.
\end{abstract}

\newpage
\section{Introduction}\news
Kinks are examples of topological solitons (for a review see \cite{MS})
in one space dimension, which interpolate between distinct vacua.
A kink can be trivially embedded into three-dimensional Euclidean space as
a solution which is independent of all but one spatial direction, and
this is known as a domain wall. The energy of a 
domain wall is proportional to its surface area, therefore the 
embedding into $\bR^3$ gives a domain wall
with infinite energy. However, the domain wall does have finite energy 
per unit area, which is the tension of the wall, and is 
equal to the kink energy in the one-dimensional theory.

Domain walls can be constructed in a three-dimensional compact space, 
in which case they have finite energy. A minimal energy configuration of 
domain walls approximates a minimal surface with increasing 
accuracy as the thickness of the domain wall is reduced. The surface
associated with a configuration of domain walls can equivalently be
defined as the level set of the field (taking the level set value to be the
midpoint between the two vacua) or as a maximal energy density isosurface.
This approach of modelling minimal surfaces using domain walls has been
successfully implemented numerically to investigate both triply periodic
 and quasicrystalline minimal surfaces \cite{GH,SE}.

In this paper, we extend the domain wall description of minimal surfaces to
double bubbles, which are global minimal area surfaces 
 which enclose and separate two prescribed volumes. We show that
intersecting domain walls in a Wess-Zumino theory with three vacua 
form double bubbles under a certain volume-preserving flow.

In 2002 it was proved \cite{proof} that in three-dimensional Euclidean space
the double bubble consists of three segments of round spheres, with 
radii $R_1\le R_2\le R_3,$ satisfying the relation 
$R_1^{-1}=R_2^{-1}+R_3^{-1},$ and meeting in triple junctions
at angles of $120^\circ.$ Such a configuration is known as the standard
double bubble and is the familiar structure seen in soap bubbles.
The proof \cite{proof} followed an earlier computer assisted proof 
\cite{comproof}, restricted to the situation in which the two
prescribed volumes are equal. 

On compact spaces the double bubble
problem is far more complicated. Even in the flat cubic three-torus,
it is still an open problem to prove the form taken by a double bubble.
The only rigorous result is that the standard double bubble is
recovered if the size of the double bubble is sufficiently small compared
to the size of the torus \cite{3dnum}. 
Numerical investigations have been performed
for the flat cubic three-torus \cite{3dnum} using Brakke's surface 
evolver software \cite{brak}, which
is based on adaptive computations of triangulations of the surface.
These numerical results suggest ten different qualitative forms taken
by the double bubble, depending upon the two bubble volumes relative
to the volume of the torus.  

In two space dimensions the double bubble problem is better understood,
and consists of determining the curve with minimal perimeter that
encloses and separates two prescribed areas. Note that in the two-dimensional
case the term volume will often be used rather than area, in order
to keep the terminology of the three-dimensional situation, but we shall
always refer to perimeter as the quantity to be minimized.
In the flat square two-torus it has been proved \cite{2d}
that the double bubble  takes one of four different qualitative forms, 
and explicit formulae are available
for the perimeter of each form as a function of the two bubble volumes.
A numerical evaluation of these formulae then allows the determination of
the form of the double bubble for any given pair of bubble volumes, and
hence the construction of a phase diagram.

To illustrate the applicability of our domain wall approach to double 
bubbles we reconstruct the phase diagram for double bubbles
in the flat square two-torus and also construct all known examples of 
double bubbles in the flat cubic three-torus.

\section{Domain walls in the Wess-Zumino model}\news
The field theory of interest in this paper is the bosonic sector
of the Wess-Zumino model, which contains a single complex scalar field
$\phi.$ For static fields the theory may be defined by its energy density,
which is given by
\be
{\cal E}=\frac{1}{4}\partial_i\phi\partial_i\bar\phi
+\left|\frac{dW}{d\phi}\right|^2,
\label{energydensity}
\ee
where $W(\phi)$ is the superpotential. For applications to double
bubbles we require a triply degenerate vacuum and therefore take
the superpotential to be
\be
W=\phi-\frac{1}{4}\phi^4.
\label{superpotential}
\ee
The zero energy vacuum field configurations are the three cube roots of
unity $\phi=\omega^j\equiv \phi_j,$ where $\omega=e^{i2\pi/3},$ with
$j\in\{1,2,3\}.$

In one space dimension this theory has six types of kink solutions (including
anti-kinks, as we make no distinction between these in this paper), which 
are the minimal energy field configurations $\phi(x)$ that interpolate
between any two distinct vacua, that is, $\phi(-\infty)=\phi_i$ and
$\phi(+\infty)=\phi_j,$ with $i\ne j.$ These kinks satisfy a 
first-order Bogomolny equation and have an energy given by
\be
E=\int^{\infty}_{-\infty}{\cal E}\,dx=|W(\phi_i)-W(\phi_j)|=3\sqrt{3}/4.
\ee
Kinks in the above theory have a width of order one, but introducing
a constant $\epsilon^2$ in front of the first term in (\ref{energydensity})
makes the kink width $O(\epsilon).$ 
The thin wall limit is $\epsilon\rightarrow 0,$ and rigorous results
require proofs based on regularity as this limit is approached.
However, as only relative scales are relevant, we find it more
convenient for a numerical implementation to fix the kink scale to be of
order one and consider all other scales in the problem (such as the
size of a region containing a single vacuum) to be $O(\epsilon^{-1}).$

A trivial embedding of kinks into a higher-dimensional version of the 
theory 
(two and three spatial dimensions will be considered) yields domain walls 
with tension $\mu=3\sqrt{3}/4.$ The tension is simply the energy per
unit area in three dimensions and the energy per unit length in two dimensions.

A kink traces out a curved path in the $\phi$-plane, connecting two
distinct vacua, but it can be shown that when viewed in the $W$-plane the
path is simply a straight line \cite{GT}. We will make use of this fact later.

This theory has a domain wall junction \cite{GT}, 
which also satisfies a first-order Bogomolny equation, 
in which the three types of domain wall mutually intersect at angles of 
$120^\circ$, 
dividing space into three equal sectors with a different vacuum value
occuring in the interior of each sector. The junction has $120^\circ$ 
angles because all domain walls have the same tension and there must be
a tension balance for an equilibrium configuration. The domain wall
junction has been computed numerically, together with networks of junctions
connected by domain wall segments that yield tilings of the plane \cite{Sa}.
 
The salient features are that the energy of a domain wall is proportional 
to its surface area, and three intersecting domain walls form a junction with 
$120^\circ$ angles. These properties suggest that domain walls 
in this system may provide a useful field theory description of double bubbles.

\section{A volume-preserving flow}\news
A bubble configuration in three dimensions consists of a spherical 
domain wall separating one vacuum field in the interior
of the bubble from a distinct vacuum field in the exterior of the bubble.
The energy of such a bubble configuration is approximately $\mu A$,
where $A$ is the surface area of the bubble and $\mu$ is the tension
introduced above.
 The error in this
 approximation can be made arbitrarily small by increasing the ratio of the 
bubble radius to the width of the domain wall, that is, by approaching 
the thin wall limit. Clearly, for the
energy density (\ref{energydensity}), there are no bubbles
that are stationary points of the energy. In order to have a bubble (and
later double bubbles) a constraint on the volume of the bubble must be
included within the field theory description, since in this theory
there is no pressure to resist the tension-induced collapse.
 In the following we describe how volume 
constraints may be included in a natural manner within the field theory.

There are two obvious ways in which dynamics may be introduced into the
static theory discussed in the previous Section. 
The first is to consider relativistic 
dynamics, which leads to the usual Lagrangian description of the 
Wess-Zumino model and yields a nonlinear wave equation which is second-order
in the time derivative. The other obvious dynamics is gradient flow,
which produces a nonlinear diffusion equation that is first-order in
time and generates an energy decreasing evolution. Given the above discussion,
it is easy to see that in both cases an
initial bubble configuration will simply collapse, as this reduces the
potential energy.

In the real Ginzburg-Landau model, it is known that
gradient flow dynamics can be modified by the introduction of 
a time-dependent effective magnetic field that preserves the global
average of the field. This volume-preserving flow is used, for example,
in the study of phase ordering and interface-controlled coarsening 
\cite{vpflow}. In this Section we describe a generalization of this
volume-preserving flow which is applicable to the Wess-Zumino model and allows
two independent volumes to be preserved during the flow.

Motivated by the applications to follow, 
we consider the theory defined on a torus $T^d,$ with volume $V.$

First of all, for each vacuum we need to introduce a density function, 
such that it is equal to unity if the field is in the given vacuum
and vanishes if the field is in any other vacuum. A simple choice is
\be
v_i(\phi)=\frac{|\phi-\phi_j|^2|\phi-\phi_k|^2}
{|\phi_i-\phi_j|^2|\phi_i-\phi_k|^2}
=\frac{1}{9}|\phi-\phi_j|^2|\phi-\phi_k|^2,
\label{vden}
\ee
where $i,j,k$ are three distinct elements of $\{1,2,3\}.$
Clearly these functions satisfy the required property that
$v_i(\phi_j)=\delta_{ij}.$ Next we define the volume occupied by
each vacuum as 
\be
V_i=\int_{T^d} v_i\, d^dx.
\label{volume}
\ee
In the thin wall limit, the width of the domain wall is negligible
compared to the volumes occupied by the vacua, and therefore in this limit  
$V_1+V_2+V_3\rightarrow V.$ 

The task is to construct an energy minimizing gradient flow in which
both volumes $V_1$ and $V_2$ remain constant.  Note that $V_3$ is
then automatically constant, at least to the accuracy in which
the thin wall limit is approached, as $V_3=V-V_1-V_2.$ 

The starting point is the gradient flow dynamics associated with the
energy density (\ref{energydensity})
\be
\frac{\partial\phi}{\partial t_0}
=-\frac{\delta {\cal E}}{\delta \bar\phi}\equiv F
=\frac{1}{4}\partial_i\partial_i\phi+3(1-\phi^3)\bar\phi^2.
\label{fflow}
\ee
The dynamics that follows from flow in the $t_0$ time variable 
reduces the energy $E$ but also changes the volumes 
$V_i.$ 

Associated with each volume density $v_i$ one can define an
alternative gradient flow 
\be
\frac{\partial\phi}{\partial t_i}
=-\frac{\partial  v_i}{\partial \bar\phi}\equiv f_i
=\frac{1}{9}(\phi-\phi_j)(\phi-\phi_k)(2\bar\phi-\bar\phi_j-\bar\phi_k),
\label{fiflow}
\ee
so that evolution in the time variable $t_i$ is a flow that
reduces the volume $V_i.$ 

The idea is to introduce a new flow, with time variable $t,$
that is constructed from the $t_0$ flow by projecting out the components
in the directions of the $t_i$ flows. This makes the $t$ flow
orthogonal to all $t_i$ flows and hence preserves all the volumes $V_i.$

The appropriate inner product is 
\be
<f,g>\,=\frac{1}{V}\int_{T^d} \bar f g \, d^dx,
\ee
and we use the notation 
\be
||f||=\sqrt{<f,f>}.
\ee
Given the flows (\ref{fiflow}) an orthonormal set of flows $\hat f_i$ is
constructed as
\be
\tilde f_i=f_i-\sum_{j<i}<\hat f_j,f_i>\hat f_j, \quad \mbox{then}\quad
\hat f_i=\frac{\tilde f_i}{||\tilde f_i||}.
\ee 
Finally, the $t$ flow is defined to be
\be
\frac{\partial\phi}{\partial t}
=F-\sum_{i=1}^2 <\hat f_i,F>\hat f_i,
\label{tflow}
\ee
which is a nonlocal and nonlinear reaction diffusion equation.
By construction the flow (\ref{tflow}) is orthogonal
to all the $t_i$ flows, that is, 
\be
<f_i,\frac{\partial \phi}{\partial t}>\,=0.
\label{orthog}
\ee

It is easy to prove that the flow (\ref{tflow}) preserves both volumes
$V_i,$ with $i=1,2$, while reducing the energy $E.$ First of all,
\be
\frac{dV_i}{dt}=\int_{T^d}\frac{dv_i}{dt}\, d^dx
=2\Re\int_{T^d}
\frac{\partial v_i}{\partial\phi}\frac{\partial \phi}{\partial t}\, d^dx
=-2\Re\int_{T^d}
\bar f_i \frac{\partial \phi}{\partial t} 
\, d^dx
=-2V\Re <f_i,\frac{\partial \phi}{\partial t}>\,=0,
\ee
where we have used the orthogonality property (\ref{orthog}).

To prove that the flow is energy decreasing, it is convenient to 
write the sum in (\ref{tflow}) as a single term, by defining
\be
\hat f=\frac{<\hat f_1,F>\hat f_1+<\hat f_2,F>\hat f_2}
{\sqrt{<\hat f_1,F>^2+<\hat f_2,F>^2}}.
\ee
Using the orthogonality property $<\hat f_1,\hat f_2>=0,$ 
allows (\ref{tflow}) to be rewritten as
\be
\frac{\partial\phi}{\partial t}
=F-<\hat f,F>\hat f.
\label{tflow2}
\ee
Therefore,
\bea
& &\frac{dE}{dt}=\int_{T^d}\frac{d {\cal E}}{dt}\, d^dx
=2\Re\int_{T^d}
\frac{\partial \bar\phi}{\partial t}
\frac{\delta {\cal E}}{\delta \bar \phi}
\, d^dx
=-2\Re\int_{T^d}
\frac{\partial \bar \phi}{\partial t} F
\, d^dx
=-2V\Re <\frac{\partial \phi}{\partial t},F>\nonumber\\
& &
=-2V\Re <\frac{\partial \phi}{\partial t},
\frac{\partial \phi}{\partial t}+<\hat f,F>\hat f>
=-2V\Re <\frac{\partial \phi}{\partial t},
\frac{\partial \phi}{\partial t}>
\,\,\le 0,
\eea
where the last expression uses the fact that the $t$ flow is
perpendicular to the combined volume reducing flow, that is,
$<\frac{\partial \phi}{\partial t},\hat f>\,=0.$

In summary, we have proved that the flow (\ref{tflow}) is energy
decreasing while preserving both volumes $V_1$ and $V_2.$
The end points of this flow are therefore equilibrium configurations
that are critical points of the energy functional under constrained
variations that preserve both volumes. Static solutions of this flow are 
candidates for double bubbles. 

The volume-preserving flow of the real Ginzburg-Landau model is 
recovered from the above generalization by restricting $\phi$ to
be real $\phi=\bar\phi\equiv \varphi$ and choosing the volume density to be 
$v_1=(\varphi-\varphi_2)/(\varphi_1-\varphi_2),$
where $\varphi_1$ and $\varphi_2=-\varphi_1$ are the two real vacua. 
In this case
$f_1=1$ and the flow takes the form
\be
\frac{\partial\varphi}{\partial t}=F-<F>,
\ee
where $<F>\,\equiv\,<1,F>$ is the average value of $F$. This flow clearly
preserves the average volume as $\frac{\partial}{\partial t}<\varphi>=0.$ 

The definition of a double
bubble requires the surface to be the global minimum of the area 
functional, with the prescribed volume constraints. Therefore, there
may be a number of surfaces which are local minima of the surface area,
or even saddle points, but are not double bubbles. All local minima
must be found to determine which is the double bubble, and this 
applies to the field theory energy computations too.

In the following two Sections we present the results of a numerical
implementation of the above volume-preserving flow. Double
bubbles in the two-torus and three-torus are studied and comparisons
made with previous results obtained using other methods that are
not based on field theory.

\section{Double bubbles in the two-torus}\news
In this Section we restrict our investigations to two-dimensional
double bubbles. As mentioned earlier, we shall retain the three-dimensional
notation and refer to the volume of a bubble (even though it is an area).
The problem is to find curves with minimal perimeter that enclose and
separate two prescribed volumes. 

Although we work on the flat square two-torus, the case of the Euclidean
plane can be recovered if the torus is sufficiently large, so that no curve of
the double bubble wraps a cycle of the torus. As a first illustration
of our method, we discuss an example of this type.
\begin{figure}[ht]
\begin{center}
\includegraphics[width=15cm]{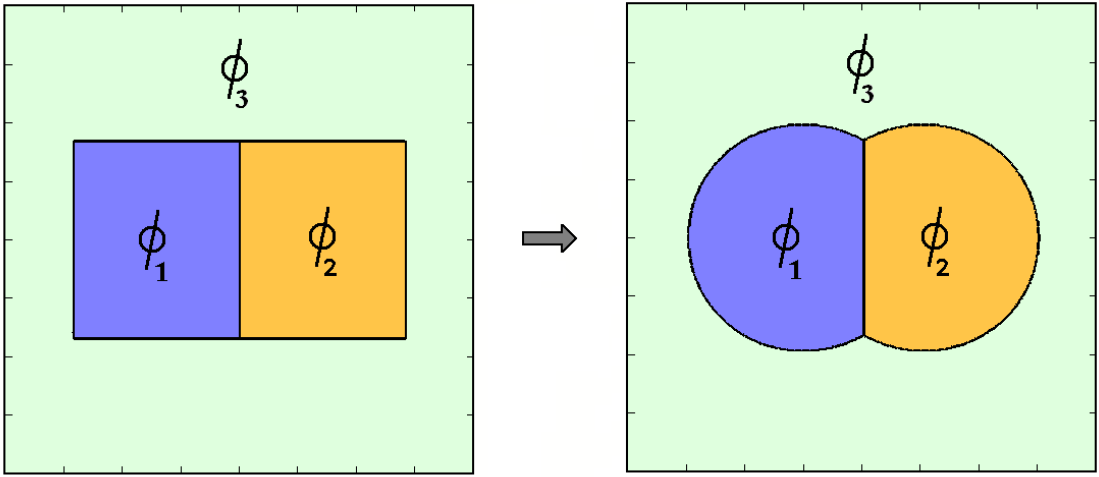}
\caption{The image on the left shows the initial condition and 
the image on the right is the subsequent final field
configuration obtained under the volume-preserving flow.
Different colours (or shades of grey) represent regions in space
where the field takes values in one of the three different vacua
$\phi_1,\phi_2,\phi_3.$ Lines denote the domain wall positions.
}
\label{fig-mg08}
\end{center}
\end{figure}

Figure \ref{fig-mg08}
 presents an initial condition and the subsequent final field
configuration obtained under the volume-preserving flow.
The lattice contains $401^2$ grid points with a lattice spacing $\Delta x=0.2$
so that the volume of the torus is $V=6400.$ The Laplacian is evaluated using
second-order accurate finite difference approximations and the flow is
evolved using a simple first-order accurate explicit method with
timestep $\Delta t=0.008.$  All inner products are evaluated by approximating 
the integral by a sum over lattice sites.

The initial condition in Figure~\ref{fig-mg08}
 consists of two equal volume rectangles, each
of volume $V_1=V_2=0.15V,$ with the field set to the vacuum value
$\phi_1$ inside one rectangle and $\phi_2$ inside
the other rectangle. Outside these rectangles the field takes
the value $\phi_3.$ The final configuration shown is at time $t=400,$
when the field is static to the accuracy that we compute. The
equilibrium field configuration clearly
takes the form of the standard double bubble in the plane with
equal volumes. Note the $120^\circ$ intersection angles of the domian
wall junctions. 

The perimeter, $P,$ of a standard double bubble in
the plane with equal volumes $V_1=V_2$ satisfies \cite{2d}
\be
\frac{P}{\sqrt{V_1}}=\frac{16\pi+6\sqrt{3}}{\sqrt{24\pi+9\sqrt{3}}}=6.359\ldots
\label{exact2d}
\ee  
Computing the perimeter length in the field theory, as the 
ratio of the energy to the tension $E/\mu,$ gives the result 
\be
\frac{E}{\mu\sqrt{V_1}}=6.370 
\label{approx2d}
\ee
therefore the field theory computation has an error of less than $\frac{1}{4}\%$
for this choice of simulation parameters, which has quite a modest grid. 
The accuracy can be improved by increasing the number of grid points, so
that the size of the torus $\sqrt{V}$ increases relative to the
width of the domain wall, that is, the system is moved closer to 
the thin wall limit. In Section~\ref{sec-mod} we discuss some 
modifications that can be applied to improve the accuracy of the computations
without the need to increase the grid size.

Note that from now on we shall refer to all volumes in units of the
torus volume, so for the above example we simply write $V_1=0.15,$ etc.
\begin{figure}[ht]
\begin{center}
\includegraphics[width=10cm]{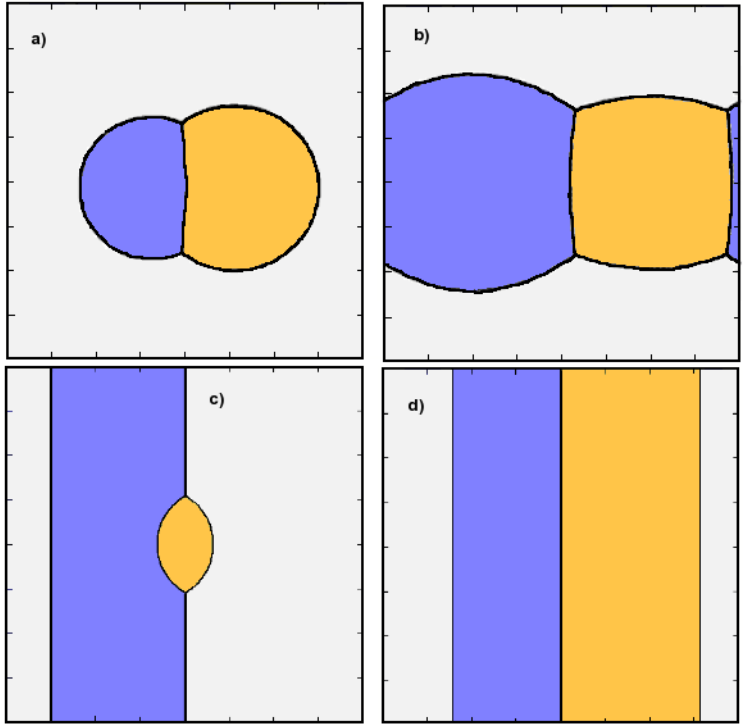}
\caption{Equilibrium fields obtained from 
four different simulations with varying volumes.
Different colours (or shades of grey) represent regions in space
where the field takes values in one of the three different vacua.
The configurations shown are known as (a) the standard double bubble,
(b) the standard chain, (c) the band lens, (d) the double band.
}
\label{fig-mg10}
\end{center}
\end{figure}    
Figure~\ref{fig-mg10} presents equilibrium fields obtained from 
four different simulations with volumes $\{V_1,V_2\}$ given by 
(a) $\{0.1,0.15\}$; (b) $\{0.3,0.2\}$; (c) $\{0.35,0.05\}$; (d) $\{0.3,0.4\}$.
In each case the initial field consists of rectangular
regions of vacuum, similar to that shown in Figure~\ref{fig-mg10}, though
some of the rectangles were chosen to wrap the torus.
In each case, for the chosen volumes, the resulting field configuration
provides a good description of the double bubble.
The configurations shown are known as (a) the standard double bubble,
(b) the standard chain, (c) the band lens, (d) the double band.
It has been proved that for all volumes $\{V_1,V_2\}$ the
double bubble takes one of these four forms, and a phase
diagram has been computed to determine which of the four forms
is taken for any given volumes \cite{2d}.

\begin{figure}[ht]
\begin{center}
\includegraphics[width=10cm]{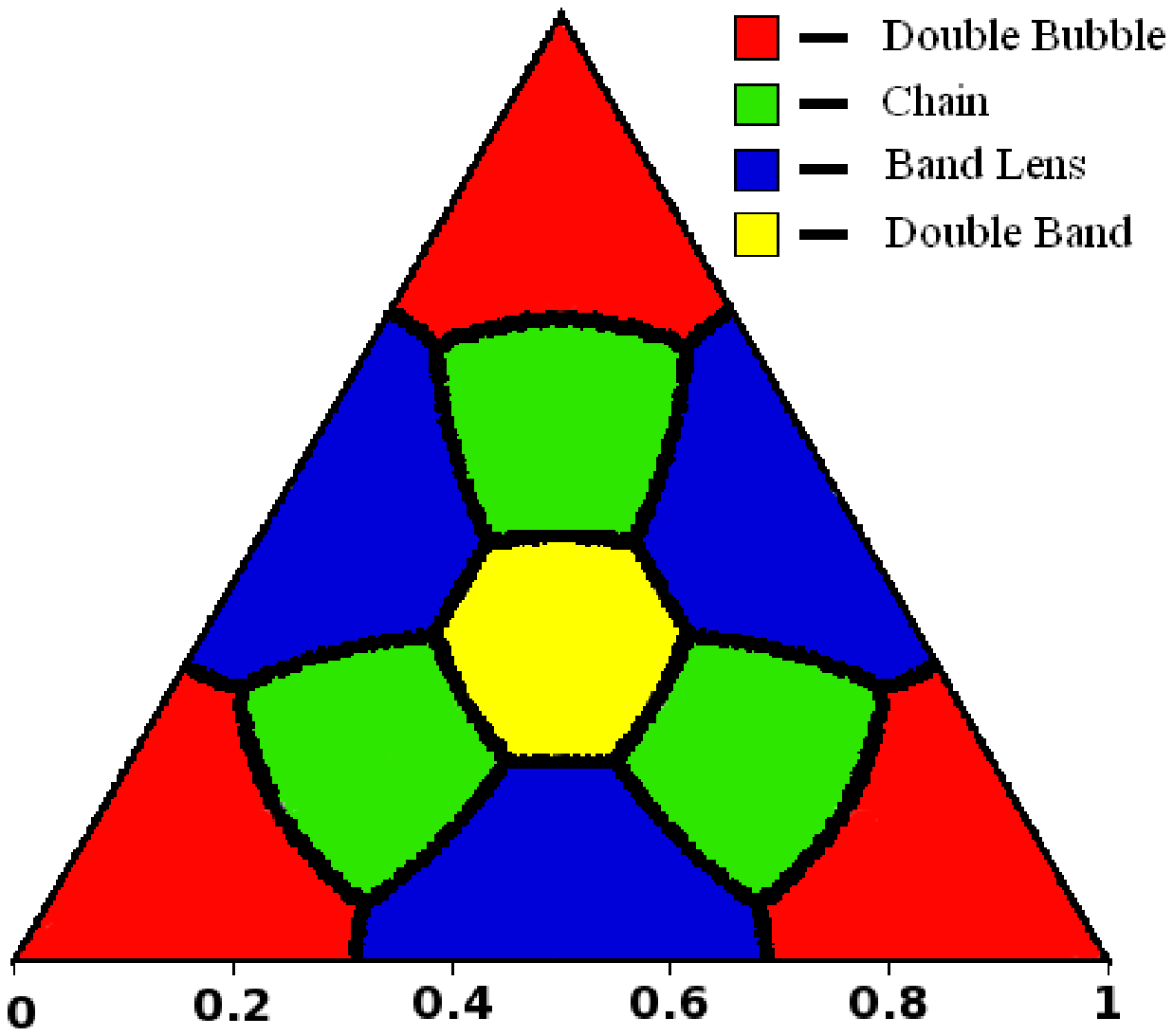}
\caption{A phase diagram indicating the form taken
by the double bubble as a function of the individual bubble volumes.
A position of a point inside the triangle is related to the
bubble volumes through equation (\ref{point}). 
The centroid of the triangle is where all three volumes are equal,
whereas regions near the vertices are where two volumes are much smaller
than the third. Finally, one of the volumes vanishes at an edge, where
the included scale gives the fraction of the total volume occupied
by one of the two remaining volumes.
}
\label{fig-mg14}
\end{center}
\end{figure}
Using the field theory description we have been able to reproduce this
phase diagram by performing simulations for a range of volumes.
This phase diagram is presented in Figure~\ref{fig-mg14}. 
The set of possible volumes,
$V_1,V_2,V_3,$ such that $V_1+V_2+V_3=1,$ is represented by the interior
of an equilateral triangle. Explicitly, the $(x,y)$ coordinates in the
plane containing the triangle are related to the volumes by
\be
(x,y)=\left(\frac{1}{2}(V_1-V_2),\frac{1}{2\sqrt{3}}(2-3V_1-3V_2)\right).
\label{point}
\ee
The centroid of the triangle is $(x,y)=(0,0)$ and corresponds to
$V_1=V_2=V_3=\frac{1}{3},$ so that all three volumes (including the volume
exterior to both bubbles) are equal. In this case, both bubble volumes
are large and therefore it is favourable to wrap both bubbles around the torus,
as this reduces perimeter length, resulting in the double band.
Note that the double bubble problem is symmetric under the interchange of any
two of the three volumes, therefore the computation can be restricted to
the region $V_3\ge V_2\ge V_1,$ and the remainder of the phase diagram
constructed from this sixth of the triangle by symmetry. 

The regions near the vertices of the triangle correspond to the situation
in which two volumes are much smaller than the third, 
and therefore the standard double bubble is 
recovered. The edges of the triangle are where one of the volumes vanishes,
with the mid-point of an edge associated with the two non-vanishing volumes
being equal. In a region around the mid-point of an edge the band lens is the
optimal form. Finally, if two volumes are reasonably similar in size, and
not too small, then the double bubble takes the form of the chain.
The phase diagram displayed in Figure~\ref{fig-mg14}
is in excellent agreement with the
one presented in \cite{2d} and confirms the applicability of our field
theory approach to double bubbles.

\section{Double bubbles in the three-torus}\news
In this Section we turn our attention to the more complicated problem of
double bubbles on the flat cubic three-torus. In this case even the
classification of the types of double bubble that exist is an open problem
and only numerical results are available \cite{3dnum}. If both bubble
volumes are small compared to the volume of the torus then the standard
double bubble in three-dimensional Euclidean space is recovered.
One also expects various three-dimensional analogues of the two-dimensional
double bubbles discussed in the previous Section. 
Numerical investigations have been performed \cite{3dnum} using 
the surface evolver software \cite{brak} and yield a number of 
forms taken by double bubbles for approriate volumes. 
These results suggest that there are ten different
forms taken by double bubbles. Equilibrium configurations were also
found that do not appear to be double bubbles for any values of
the volumes.

\begin{figure}
\begin{center}
\includegraphics[width=17cm]{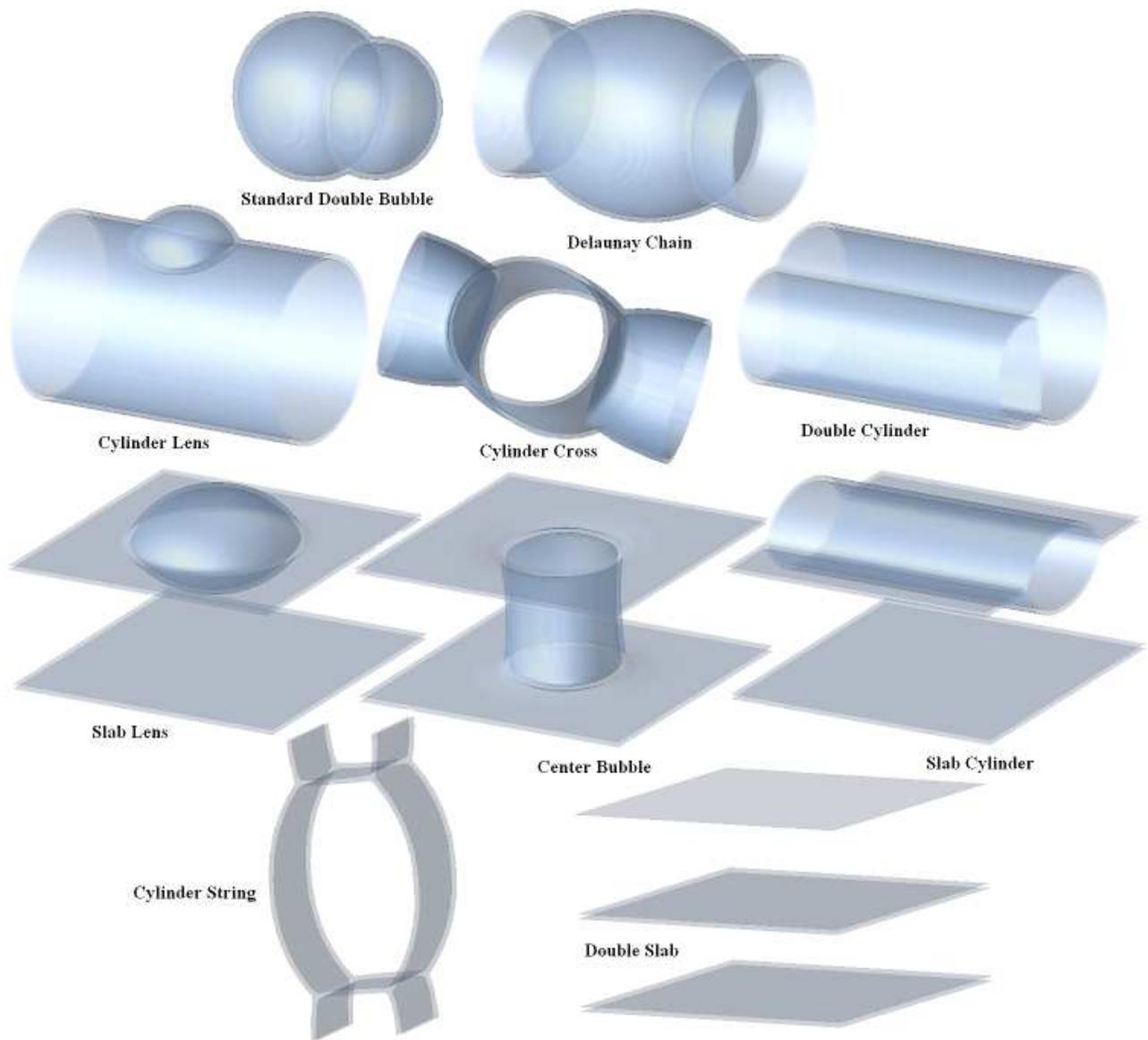}
\caption{Energy density isosurfaces for equilibrium 
configurations corresponding to the
ten different conjectured double bubble forms.
}
\label{fig-mg16}
\end{center}
\end{figure}        
Using our field theory approach with domain walls, which we stress is
very different to the surface triangulation method used in \cite{3dnum},
we have been able to reproduce all ten types of equilibrium
configurations conjectured to be doubles bubbles in \cite{3dnum}.
In Figure~\ref{fig-mg16} 
we display the ten different conjectured double bubble forms, by
plotting an energy density isosurface for each, and we include the name
of each surface using the nomenclature of \cite{3dnum}.

The initial conditions used to generate these surfaces were
constructed using the same approach as in the two-dimensional
case described in the previous Section,
namely, we take 
a collection of cuboids with assigned vacuum fields.

\begin{figure}[ht]
\begin{center}
\includegraphics[width=15cm]{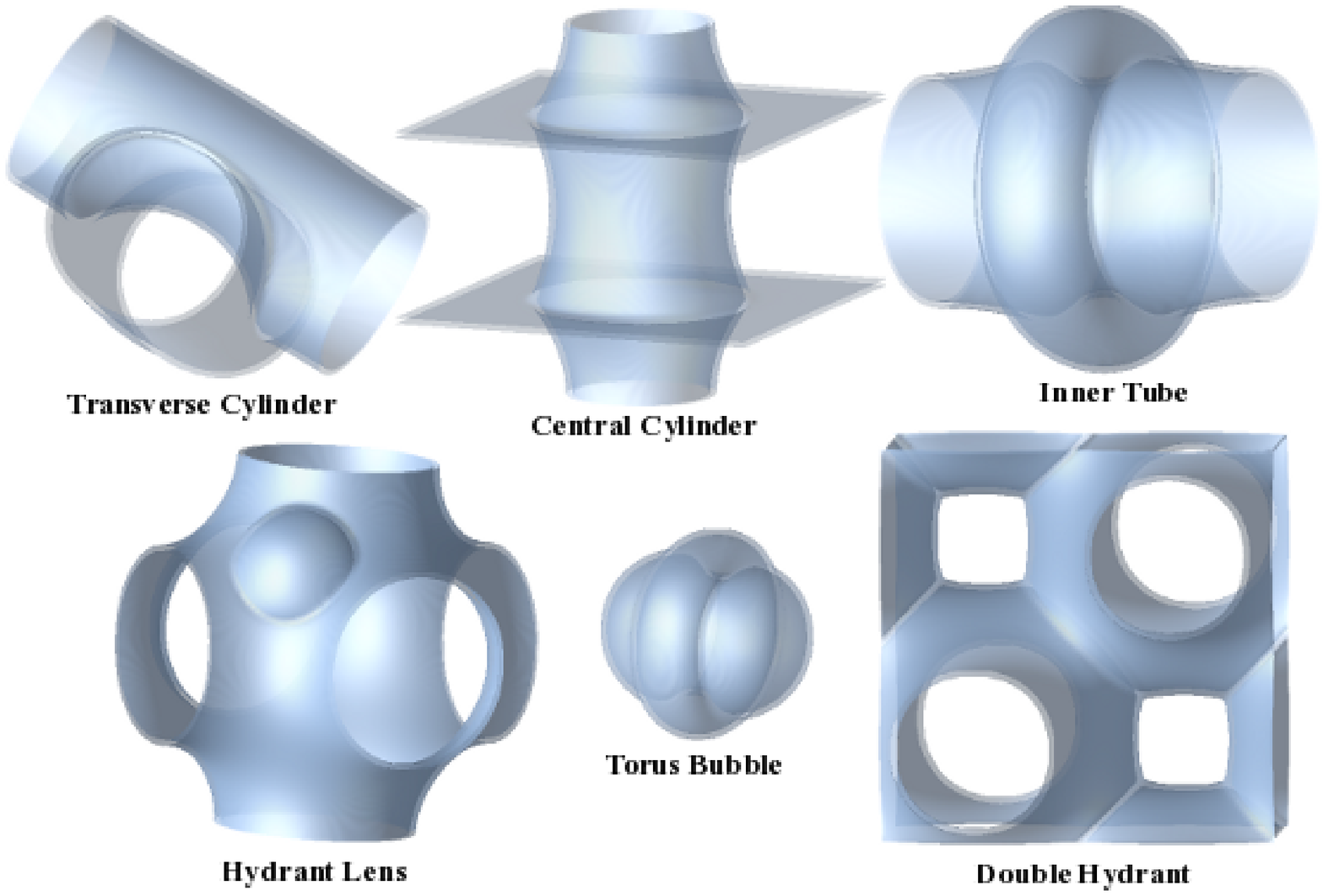}
\caption{Energy density isosurfaces for six
types of equilibrium configurations which are never double
bubbles, that is, none is the global minimal energy surface for
any values of the volumes.
}
\label{fig-mg17}
\end{center}
\end{figure}
Six types of equilibrium configurations are presented
in Figure~\ref{fig-mg17} which are never double
bubbles, that is, none is the global minimal energy surface for
any values of the volumes.
The transverse cylinder, torus bubble, inner tube
and double hydrant are saddle points
which are unstable to symmetry breaking perturbations. 
The central cylinder and hydrant lens appear
to be local minima, for suitable values of the volumes.
\begin{figure}
\begin{center}
\includegraphics[width=9cm]{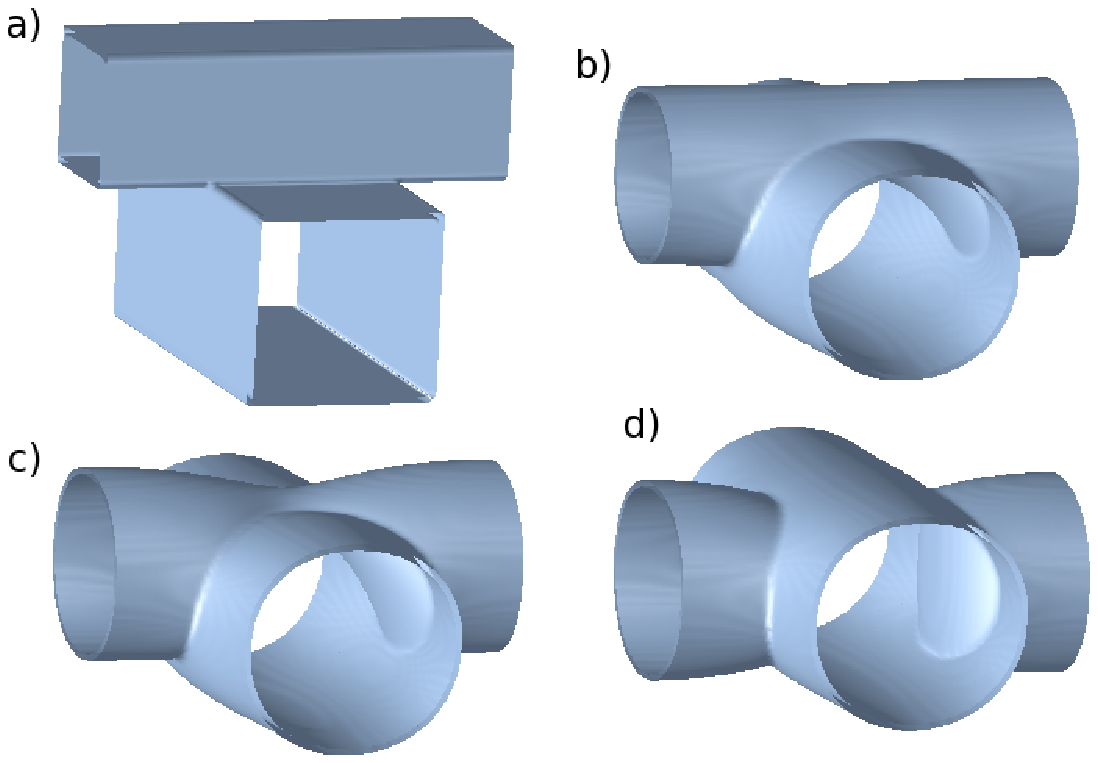}
\caption{Energy density isosurfaces at increasing times under
volume-preserving flow. The initial condition 
is of a similar type to the transverse
cylinder but evolves into a cylinder cross. 
}
\label{fig-mgtc}
\end{center}
\end{figure}
An example evolution under volume-preserving flow is presented in 
Figure~\ref{fig-mgtc}.
The initial condition 
Figure~\ref{fig-mgtc}(a) is of a similar type to the transverse
cylinder, but under the evolution it evolves into a 
stable cylinder cross Figure~\ref{fig-mgtc}(d).
This simulation demonstrates both the instability of the transverse
cylinder and the stability of the cylinder cross.

The results presented in this Section confirm the applicability of the
domain wall approach to three-dimensional double bubbles and show that the
method can be numerically realized with reasonable computing resources.
Evolutions that involve topology changes of the surface are automatically
dealt with within the field theory description, since the surface is obtained
as a level set. This allows a large region of configuration space to be 
explored without the need for fine tuned initial conditions, as demonstrated
in Figure~\ref{fig-mgtc}, where the final equilibrium configuration 
is quite different from the initial condition. 
Our results provide some confirmation of the numerical
results in \cite{3dnum} using very different methods, and therefore provides
evidence for the accuracy of both approaches. The phase diagram for 
three-dimensional double bubbles computed in \cite{3dnum} could also be
obtained using our field theory approach, though we have not pursued
that here. 

\section{Modified volumes and penalty functions}\label{sec-mod}\news
In this Section we discuss an improvement on the simple volume 
density functions introduced earlier and also 
describe an alternative method to the volume-preserving flow.

The volume density functions (\ref{vden}) satisfy the minimal
requirement that $v_i(\phi_j)=\delta_{ij}.$ In the thin wall limit
the volume of any region in which the field is not in a vacuum value
tends to zero, so the above minimal requirement is all that is relevant.
However, in any numerical implementation there is a 
non-zero value of the wall thickness 
(that is, $\epsilon\ne 0$), and the volume density (\ref{vden}) 
will provide a contribution
to the volume as the domain wall is crossed. Considering the volume $V_i$ then
the density will be non-zero in the interior of the domain wall
connecting vacuum $\phi_i$ to $\phi_j.$ This is reasonable since the
region containing vacuum $\phi_i$ now has a diffuse rather than sharp
boundary. However, the simple volume density (\ref{vden}) is also
non-zero in the interior of the domain wall connecting vacuum $\phi_j$
to $\phi_k$ and this is certainly not related to the volume $V_i.$
Of course, as the thin wall limit is approached this effect tends to
zero, but from the point of view of improving numerical accuracy it
would be desirable if this contribution could be eliminated for any 
wall thickness. This can be achieved using the modification presented below.
    
The above comments lead to an additional requirement on the volume
density, namely that $v_i(\phi_{jk})=0,$ where $\phi_{jk}$ denotes the
domain wall solution connecting vacuum $\phi_j$ to $\phi_k.$
Recall that although a domain wall solution traces out a curved path
in the $\phi$-plane it is a straight line in 
the $W$-plane. The additional requirement
can therefore be met by choosing a volume density $v_i(\phi)$ that 
vanishes along
the whole line in the $W$-plane that connects the points $W(\phi_j)$ and
$W(\phi_k).$ The simplest choice is to take
\be
v_i(\phi)=
\bigg(\frac{
\Im\{(W(\phi)-W(\phi_k))(\overline{W(\phi_j)}-\overline{W(\phi_k)})\}}
{\Im\{(W(\phi_i)-W(\phi_k))(\overline{W(\phi_j)}-\overline{W(\phi_k)})\}}
\bigg)^2,
\label{modvden}
\ee
which clearly satisfies the required properties.

We have used this modified definition of the volume density and found that,
to a good accuracy, the same results are obtained as with (\ref{vden}).
The modified volume density indeed solves the problem of spurious 
contributions to the volume, though as mentioned earlier, such errors 
were already small by virtue of the fact that they vanish as the 
thin wall limit is approached.

The volume-preserving flow introduced in this paper is an elegant 
method to minimize energy while constraining volumes. A less
sophisticated approach, which is often used in constrained numerical
minimization, is to introduce a penalty function. Explicitly, this 
involves the addition to the field theory energy of a contribution of the
form
\be
E_{penalty}=\lambda\{(V_1-V_1^*)^2+(V_2-V_2^*)^2\},
\ee
where $V_1^*$ and $V_2^*$ are the two required values of the 
volumes $V_1$ and $V_2$ and $\lambda$ is a large positive constant,
$\lambda \gg 1.$ 

This additional contribution to the energy is known as a penalty function,
since it penalizes field configurations that do not have the required 
values for the volumes.
In the limit as $\lambda\rightarrow\infty$ the penalty
function enforces the required volumes $V_i=V_i^*.$ 
If standard gradient flow is applied to the energy function with the
additional term and a finite value of $\lambda$ then the volumes will
approach the required values, to an arbitrary accuracy controlled
by the value $\lambda.$ For fields with the correct volumes then the 
additional contribution to the energy vanishes and the usual energy 
then dominates, producing field configurations that minimize the 
usual energy with the given fixed volumes as a constraint.

The above penalty function method is not as elegant as the
volume-preserving flow technique and requires a careful choice of $\lambda,$
to ensure that the volume constraints are satisfied to a good accuracy
whilst preventing the penalty contribution from completely dominating
over the remaining energy term. However, it does have one 
practical advantage concerning global minima and equilibrium 
configurations, as we now discuss.

It is easy to see that there are equilibrium surfaces, that is, stationary
points of the area functional restricted to volume-preserving variations,
which have non-trivial zero modes, 
and therefore are not even local energy minima.
A simple example in $\bR^3$ is one spherical bubble entirely 
contained within a 
second spherical bubble. This is an equilibrium configuration which has a zero
mode corresponding to the translation of the inner bubble, so that it remains
entirely within the outer bubble. The corresponding field theory
configuration is also an equilibrium solution and the volume-preserving
flow will end at such a static solution given appropriate initial conditions.
To find the global energy minimum, which is the standard double bubble, 
requires an appropriate choice of initial conditions to avoid the flow
getting trapped in an equilibrium solution of the above type. 
More complicated examples of the above phenomenon also exist, such as 
equilibrium chain configuration in $T^2,$ where each junction is in
equilibrium but the relative proportions of the chain segments are not those
of minimal area. 

Even in the thin wall limit, the penalty function method does not solve 
precisely the same problem as area minimization for given fixed volumes,
because $\lambda$ is finite. This difference is sufficient to remove the 
zero-modes discussed above and it turns out that a numerical implementation
of the penalty function method yields global energy minima from a much
larger set of initial conditions than the volume-preserving flow technique.
In fact, the penalty function method was used to construct the 
double bubble chains in two and three dimensions presented earlier, 
as it is more efficient than using volume-preserving flow to compute
solutions of the chain type.

\section{Conclusion}\news
We have presented a field theory description of double bubbles using
domain walls in a Wess-Zumino model with a volume-preserving flow.
The applicability of this approach has been demonstrated by reproducing
the phase diagram for double bubbles in the flat square two-torus and all 
examples for candidate double bubbles in the flat cubic three-torus.

In addition to providing an alternative numerical approach to computing
double bubbles one may speculate that the field theory formulation
in terms of a volume-preserving gradient flow might be useful in proving 
rigorous mathematical results, along the same lines that Ricci flow 
has proved to be such a great tool in the study of surfaces. 
 
By using a higher-order superpotential, a field theory with
more than three vacua can be constructed and the volume-preserving
flow can be used to find equilibrium configurations. 
However, such a system does not describe triple bubbles, or their
multi-bubble generalizations, since not all domain wall tensions will be
equal. To provide a field theory description of multi-bubbles
requires additional fields beyond that of a single
complex scalar, so that more than three vacua can exist with
all domain wall tensions being equal. For example, for triple
bubbles an appropriate field would be a real three-component field
with four vacua and a tetrahedrally invariant potential.

\section*{Acknowledgements}
MG thanks the EPSRC for a research studentship.
PMS thanks Gary Gibbons and Stephen Watson for useful discussions.
The numerical computations were performed on the Durham HPC cluster HAMILTON.

\end{document}